\documentclass[twocolumn,pre,english,amsmath,amssymb,superscriptaddress,aps,longbibliography]{revtex4-1}
\usepackage{latexsym}
\usepackage{epsfig,psfrag}
\usepackage{dcolumn}
\usepackage{bm}
\usepackage{graphicx}
\usepackage{color}
\usepackage{esint}
\usepackage{babel}
\usepackage{amsfonts}
\usepackage{enumerate}
\usepackage{mathbbol}
\usepackage[usenames,dvipsnames]{xcolor}
\usepackage[colorlinks=true ,citecolor=blue]{hyperref}
\usepackage{lineno}

\begin{document}
\title{Spin-orbit interaction and snake states in a graphene \emph{p-n} junction}
\author{Dario Bercioux}
\email{dario.bercioux@dipc.org}
\affiliation{Donostia International Physics Center (DIPC), Manuel de Lardizbal 4, E-20018 San Sebasti\'an, Spain}
\affiliation{IKERBASQUE, Basque Foundation for Science, Maria Diaz de Haro 3, 48013 Bilbao, Spain}

\author{Alessandro De Martino}
\email{ademarti@city.ac.uk}
\affiliation{Department of Mathematics, City, University of London, London EC1V 0HB, United Kingdom}

\date{\today}

\begin{abstract}
We study a model of a \textit{p-n} junction in single-layer graphene in the presence of a perpendicular magnetic field
and spin-orbit interactions. By solving the relevant quantum-mechanical problem for a potential step, we determine the exact spectrum of 
spin-resolved dispersive Landau levels. Close to zero energy, we find a pair of linearly dispersing zero modes, 
which possess a wave-vector-dependent spin polarization and can be regarded as quantum analogs of spinful snake states.
We show that the Rashba spin-orbit interaction, in particular, produces a wave vector shift between the dispersions of these modes
with observable interference effects. These effects can in principle provide a way to detect the presence of Rashba spin-orbit interaction
and measure its strength. Our results suggest that a graphene \textit{p-n}  junction in the presence of strong
spin-orbit interaction  could be used as a building block in a spin field-effect transistor. 

\end{abstract}
 
\maketitle


\section{Introduction}

Fifteen years after the isolation of graphene~\cite{Novoselov:2004,Novoselov:2005}, 
fundamental research on this exceptional two-dimensional material is still very active.
Spin-orbit interaction (SOI) effects, in particular, continue to attract much attention.
Graphene is, in fact, a promising material for spintronic applications~\cite{Zutic:2004,Han:2014dc,Bercioux:2015,Offidani:2017km}, 
where control of the SOI plays a crucial role. Moreover, SOI is at the basis of fundamental phenomena  
as the quantum spin Hall effect, whose theoretical discovery in graphene~\cite{Kane:2005} started the field of 
topological matter~\cite{Hasan:2010ku,Chiu:2016et,Bercioux:2018}.

Due to the small atomic number of carbon, the SOI in pristine graphene is very weak (on the order of  $10\,\mu$eV)
~\cite{Gmitra:2009fh,Konschuh:2010cy}. 
However, in the last few years there has been an increasing body of experimental 
and theoretical evidence that SOI can be artificially enhanced either by chemical functionalization 
or by placing graphene on strong-SOI substrates~\cite{CastroNeto:2009bg,Weeks:2011cu,Hu:2012iq,
Marchenko:2012,Balakrishnan:2013gm,Balakrishnan:2014jg,Otrokov:2018}.
The latter approach is particularly  promising, as it preserves the electronic properties of graphene. 
While the first experimental reports presented signatures of enhanced SOI only in the spectral properties,
a breakthrough came with the advent of two-dimensional van der Waals materials~\cite{Geim:2013hf}, 
in particular the transition metal dichalcogenides (TMDC)  WSe$_2$, WS$_2$, and MoS$_2$~\cite{Roldan:2014}.  
Graphene/TMDC heterostructures allow for high quality interfaces,  
which preserve the large mobility of graphene and  provide ideal platforms for transport measurements,
showing  uniform enhancement of SOI of up to three orders of magnitude~\cite{Avsar:2014ex,Gmitra:2015kg,Gmitra:2016fk,
Wang:2015,Wang:2016,Yang:2017ds,Benitez:2017kh,Cummings:2017kb,Wakamura:2018a,Zihlmann:2018,Wakamura:2019ca,Garcia:2018kb}.
For example, the Rashba SOI for graphene on WS$_2$ is estimated to have a magnitude of up to $5$~meV~\cite{Wang:2015,Wang:2016},
and even larger for graphene on a metallic substrate, where it can reach a value of the 
order of $50$~meV~\cite{Otrokov:2018,Marchenko:2012}.

Motivated by these developments, we study here the effects of strong SOI on the 
electronic properties of graphene \textit{p-n} junctions in a perpendicular magnetic field. 
From the early days of graphene research, these systems have attracted much attention, 
in part due to the peculiar transport channels that form at the interface between the $p$- and $n$-doped 
regions~\cite{Williams:2007dr,Abanin:2007,Ozyilmaz:2007fj}. 
These chirally propagating electronic states can be visualized as {\em snake states}, whose quasiclassical trajectories describe a 
snaking motion along the interface. They play an important role in  the magnetotransport properties of these 
devices~\cite{Williams:2011kb,Barbier:2012dj,Liu:2015exa,Taychatanapat:2015ff,Rickhaus:2015cpa,Cohnitz:2016,Brouwer:2016,
Tovari:2016dw,Kolasinski:2017js,Makk:2018}. 
Similar snake states are expected also for inhomogeneous magnetic fields containing a
line separating $B>0$ and $B<0$ regions,  without a $p$-$n$ junction \cite{Park:2008gw,Ghosh:2008bo,Oroszlany:2008dn}.
These purely magnetic snake states might give rise to effects analogous to those presented in this paper, 
but we do not further consider them here. We note in passing that the effects of SOI on the  transport properties  of \textit{p-n} junctions in the absence of 
magnetic field have been studied in~\cite{Yamakage:2009}.

In this article, we investigate how the presence of finite SOI changes the interface states in a graphene \textit{p-n} junction. 
We focus on the low-energy physics described by the Dirac-Weyl Hamiltonian, 
and include in our model two SOI terms~\footnote{Recently, a third type of SOI has been argued to exist in graphene/TMDC heterostructures, 
the so-called valley-Zeeman SOI~\cite{Gmitra:2015kg,Wang:2015,Wang:2016,Gmitra:2016fk,Cummings:2017kb,Zihlmann:2018}.
This term has the form of a Zeeman coupling to an 
effective perpendicular magnetic field having opposite sign at the two valleys. 
Thus, within the single-valley description we use in this paper, one could in principle incorporate it 
in the Zeeman coupling.}: the Rashba SOI~\cite{Kane:2005,Rashba:2009}, due to inversion symmetry breaking by the
substrate or, possibly,  by a perpendicular electric field; and the Kane and Mele intrinsic SOI~\cite{Kane:2005}.
We present the exact solution of the corresponding quantum mechanical problem 
in the presence of a potential step, and thereby provide a full description of the energy 
spectrum and the eigenstates of the system. We find infinitely many branches of
dispersive Landau levels (LLs), with flat portions separated by regions with approximately linear dispersion.
The shape of the levels can be understood as follows. The electron states are parameterized by the wave 
vector along the junction, $k$, which defines the centre around which the wave function is 
localized within a magnetic length. At large $k$, the states are localized deep in the bulk, away from the 
junction, and their energy is essentially independent of $k$. These states correspond to graphene's LLs modified by the SOI. 
At small $k$, the states are localized close to the interface and feel the potential step.
Their energy acquires a $k$-dependence and, as a consequence, the corresponding states have a finite 
group velocity along the junction. Due to the presence of the magnetic field, the system is chiral and 
the group velocity has the same sign for all branches.
We will focus in particular on the two chiral modes evolving from the pristine 
zero-energy LLs.
These modes are responsible for the low-energy transport properties when the Fermi level is close to the charge neutrality point. 
We refer to them as the snake modes.   We will show that their group velocity depends on the size of the potential step, it is 
proportional to the classical drift velocity in crossed electric and magnetic fields, and it is only weakly
affected by the SOI. 

In the absence of SOI and Zeeman coupling, the snake modes are degenerate. 
We show that the Rashba SOI removes this degeneracy and leads to a relative wave vector shift between their dispersions. 
This resembles the relative horizontal shift of the energy dispersions of 
spin-up and spin-down electrons which occurs in a single-channel quantum wire in the presence of Rashba SOI, and suggests that 
graphene  \textit{p-n} junctions could be exploited to realize the Datta-Das effect~\cite{Datta:1990,Bercioux:2015}.
Indeed, electrons injected at a given energy  at some point along the junction with a definite spin polarization
will populate the two modes and will acquire different phase shifts while propagating. Monitoring the current downstream,
one would observe a modulation due to quantum interference effects between the two modes.

Before closing this section, let us briefly describe how this paper is organized.
In Sec.~\ref{secII} we formulate the model for a graphene \textit{p-n} junction 
in a perpendicular magnetic field  in the presence of finite SOIs, and review its exact solution for a uniform potential.  
In Sec.~\ref{secIII} we determine the exact spectrum of eigenstates of the quantum mechanical problem with the potential step. 
In Sec.~\ref{secIV} we discuss the corresponding observables, namely group velocity and spin polarization.
Finally, in Sec.~\ref{secV} we present some conclusions and discuss possible extensions of our work. Some technical details 
are relegated to the appendix.

\section{Model and general solution}
\label{secII}
%
%
\begin{figure}[!tb]
\includegraphics[width=0.7\columnwidth]{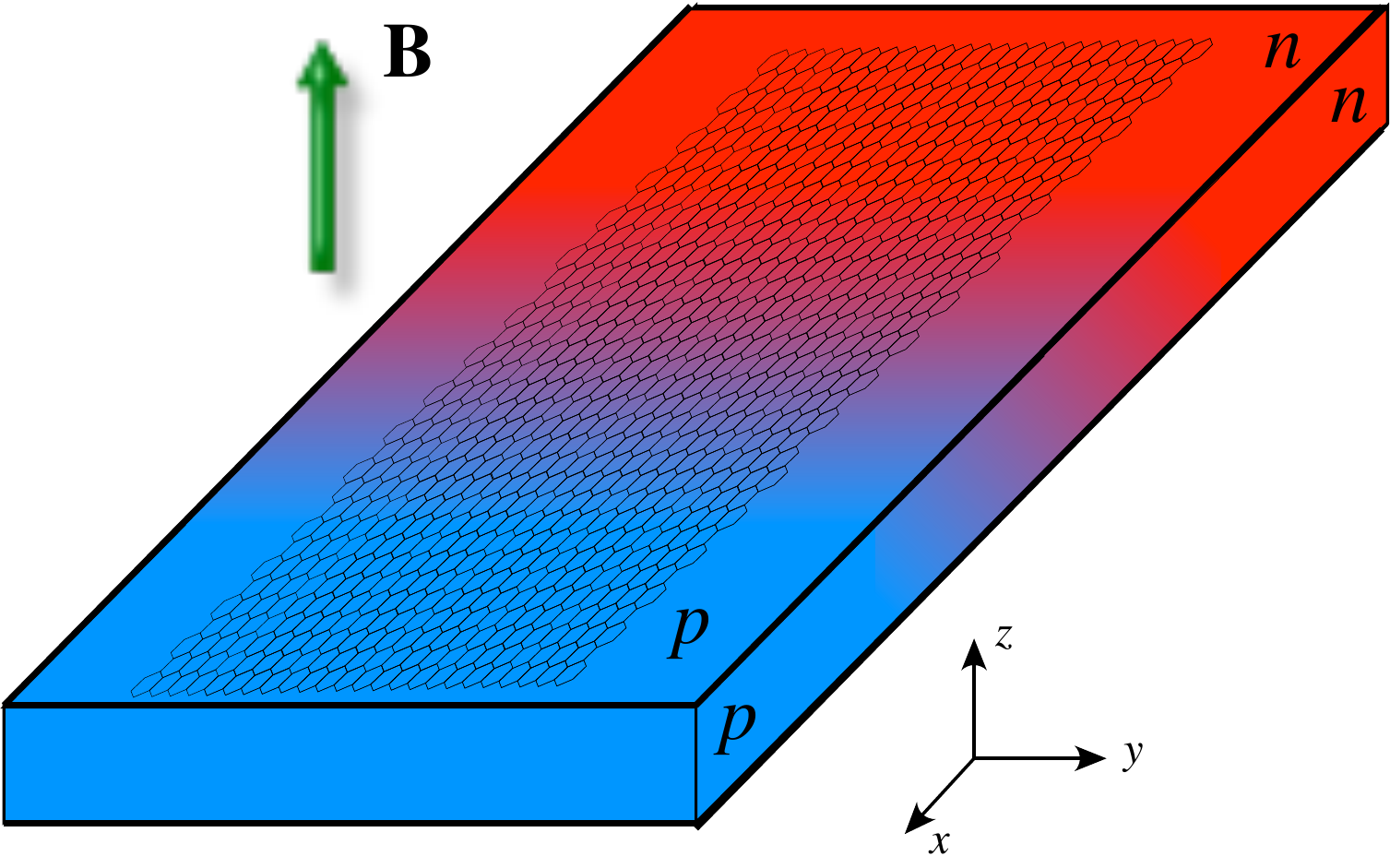}
\caption{Sketch of a graphene \textit{p-n} junction on a substrate inducing a
strong SOI in the graphene layer  by proximity effect. The length scale over which the applied potential
changes polarity is assumed to be much larger than graphene's lattice constant $a_0$.}
\label{fig1} 
\end{figure}
%
%
Our starting point is the Dirac-Weyl Hamiltonian 
for graphene in a perpendicular magnetic field ${\bf B}=B\hat z$ with SOI 
~\cite{CastroNeto:2009,Rashba:2009,Bercioux:2010,Liu:2012}:
%
%
\begin{align}
\mathcal{H}_\tau\!&=\! v_\text{F}\!  \left( \tau \sigma_x  \Pi_x +  \sigma_y  \Pi_y \right)\!+ V\!\!+
\mathcal{H}_\text{Z}+ \mathcal{H}_\text{RSO} + \mathcal{H}_\text{ISO} ,
\label{ham}
\end{align}
%
%
where $\bm{\Pi}  = \textbf{p} + \frac{e}{c} \textbf{A}$, 
 $\textbf{A}$ and $V$ are respectively vector and scalar potential,
and ${\mathcal H}_\text{Z}$ is the Zeeman term
%
%
\begin{align}
\mathcal{H}_\text{Z} = \frac{g_s\mu_B }{2}B  s_z\, .
\end{align}
%
%
We include both the Rashba SOI and the Kane-Mele intrinsic  SOI:
%
%
\begin{subequations}\label{SOI}
\begin{align}
\mathcal{H}_\text{RSO} & = \frac{\lambda}{2} \left( \tau \sigma_x s_y - \sigma_y s_x \right) \,,\label{RSOI}  \\ 
\mathcal{H}_\text{ISO} & =    \Delta( \tau\sigma_z s_z)\, . \label{ISOI}   
\end{align}
\end{subequations}
%
%
The symbol  ${\bm \sigma}=(\sigma_x,\sigma_y,\sigma_z)$ (resp. ${\bf s}=(s_x,s_y,s_z)$) 
denotes as usual the Pauli matrices in sublattice (resp. spin) space, while $\tau=\pm$ refers to the valleys $K/K'$.
In Eq.~\eqref{ham} we have assumed that the two valleys are decoupled, which holds provided 
all potentials are smooth on the scale of graphene's lattice constant~\footnote{Our analysis does not apply to situations 
in which the inter-valley coupling might play an important role as, for instance, in the presence of short-range impurities or defects.}.
The Hamiltonians at $K$ and $K'$ are related by a unitary transformation,
%
%
\begin{align}
\mathcal{H}_- = (i\sigma_y) \mathcal{H}_+ (-i\sigma_y), \label{unitarytransf}
\end{align}
%
%
which amounts to a change of sublattice basis at the valley $K'$. 
Equation~\eqref{unitarytransf} implies that the spectra of eigenvalues 
of ${\mathcal H}_+$ and ${\mathcal H}_-$ are the same 
and  that the eigenstates at $K'$ are obtained from the eigenstates at $K$ by multiplication by $i\sigma_y$.
This mapping does not affect any of the observables discussed below.
We can then restrict ourselves to a single valley, say $\tau=+1$, and 
suppress the valley index from now on.

Throughout this paper we shall use dimensionless variables, where lengths are expressed
in units of magnetic length $\ell_B=\sqrt{\hbar c/eB}$, wave vectors in units of $\ell_B^{-1}$, 
and energies in units of the relativistic cyclotron energy $\hbar \omega_c=\hbar v_\text{F}/\ell_B$.
For a magnetic field of $2$\,T, one finds $\ell_B \simeq  18$\,nm and $\hbar \omega_c \simeq 34$\,meV,  
while the Zeeman coupling strength is approximately $100\, \mu$eV.

We shall study a linear geometry, see the sketch in Fig.~\ref{fig1}, where the potential profile
is a sharp antisymmetric step,
%
%
\begin{align}
V(x)=V_0 \, \text{sign}(x), \quad V_0>0.
\label{step}
\end{align} 
%
%
The potential creates an interface at $x=0$ separating a $n$-doped region for $x<0$ from 
a $p$-doped region for $x>0$. The local in-plane electric field due to the potential step might 
in principle induce an additional local SOI, but  this is negligible due to the assumed 
smoothness of the electrostatic potential on the scale of the lattice constant~\footnote{To estimate the effect, we follow Ref.~\cite{Kane:2005} and replace the Coulomb potential with the potential of the \textit{p-n} junction. Assuming that the potential difference at the junction is of the order of the cyclotron energy and that the potential varies on the scale of the magnetic length, we estimate the induced SOI to be of the order of  $10^{-6}$~meV.}.

In the Landau gauge $\textbf{A}=(0,Bx,0)$ the Hamiltonian~\eqref{ham} is translationally 
invariant in the $y$-direction, and its eigenstates take the form
$$
\Psi_k(x,y) = \frac{1}{\sqrt{L}}e^{i k y}\psi_k(x),
$$ 
where $\psi_k(x)$ is a four component wave function describing the 
electron amplitudes on  A/B sublattices for spin up/down:
%
%
\begin{equation}
\psi_k^T(x) = \left( \psi_{\text{A}\uparrow},\psi_{\text{B}\uparrow},\psi_{\text{A}\downarrow},
\psi_{\text{B}\downarrow}\right).
\end{equation}
%
%
(Here, $T$ denotes transposition and, from now on, we shall omit
the index $k$ on the spinors.) After projection onto the plane waves in 
the $y$-direction, we need to solve the equation
\begin{equation}
\left[ {\mathcal H} (k) - E \right] \psi(x) =0,
\label{weyleq}
\end{equation}
where ${\cal H}(k)=e^{-iky} {\mathcal H}e^{iky}$. It is convenient
to introduce the ladder operators
%
%
\begin{align}
\hat a=\partial_q +\frac{q}{2}, \qquad \hat a^\dagger= - \partial_q +\frac{q}{2},
\qquad [a,a^\dagger] =1,
\end{align}
%
%
where $q=\sqrt{2}(x + k)$ is the shifted and rescaled spatial coordinate
and  $\partial_q= \frac{1}{\sqrt{2}}\partial_x$ the corresponding differential operator.
Then we have
%
%
\begin{align}\label{modelham}
{\mathcal H}(k) -E= 
\begin{pmatrix}
-\mu_+ & - i\sqrt{2}\hat a& 0 & 0 \\
 i\sqrt{2}\hat a^\dagger& -\mu_-  & -i\lambda & 0 \\
0 & i\lambda & -\nu_- & -i\sqrt{2}\hat a\\
0 & 0 & i \sqrt{2}\hat a^\dagger& -\nu_+
\end{pmatrix},
\end{align}
%
%
where we use the notation  ($\alpha=\pm $)
%
%
\begin{align*}
\mu_\alpha& =E-(V+b+\alpha\Delta), \\
\nu_\alpha& = E-(V-b+\alpha\Delta) ,
\end{align*}
%
%
and $b$ denotes the dimensionless Zeeman coupling. 

The general solution to Eq.~\eqref{weyleq} for piecewise constant external fields
was obtained in~\cite{DeMartino:2011}
(see also~\cite{Rashba:2009}).
To make this paper self-contained, we concisely review the main steps of the solution. 
The structure of Eq.~\eqref{modelham} suggests the following ansatz:
%
%
\begin{align}
\psi^{\gtrless}(x)
=
\begin{pmatrix}
d_1 D_{p-2}(\pm q) \\ i d_2 D_{p-1}(\pm q) \\ d_3 D_{p-1}(\pm q) \\ i d_4 D_{p}(\pm q)
\end{pmatrix},
\label{wf}
\end{align}
%
%
where $D_p(\pm q)$ are the parabolic cylinder functions~\cite{Olver:2010} and $p$ is a real
parameter. The parabolic cylinder functions are the eigenfunctions of the operator
$\hat a^\dagger \hat a$, 
$
\hat a^\dagger \hat aD_p(\pm q) =p D_p(\pm  q)$, 
and satisfy the following recurrence relations:
%
%
\begin{align}
\hat aD_p(\pm q) &=\pm pD_{p-1}(\pm q), \nonumber \\
\hat a^\dagger D_p(\pm q) & = \pm D_{p+1}(\pm q). \nonumber
\end{align}
%
%
The superscript $>$ (resp. $<$) in Eq.~\eqref{wf} indicates that the wave function vanishes for 
$x \rightarrow + \infty$ (resp. $x \rightarrow - \infty$).
Inserting this ansatz into Eq.~\eqref{weyleq}, we get a
homogeneous  linear system of equations for the coefficients $\{d_i\}$:
%
%
\begin{equation}
\begin{pmatrix}
-\mu_+ &  \pm \sqrt{2}(p-1) & 0 & 0 \\
\pm \sqrt{2} & -\mu_-  & -\lambda & 0 \\
0 & - \lambda &- \nu_- & \pm \sqrt{2} p \\
0 & 0 &  \pm  \sqrt{2} & - \nu_+
\end{pmatrix}
\begin{pmatrix}
d_1\\
d_2\\
d_3\\
d_4
\end{pmatrix}
=  \begin{pmatrix}
0\\
0\\
0\\
0
\end{pmatrix}.
\label{dcoeff}
\end{equation}
%
%
The existence of a non-trivial solution requires  the vanishing of the determinant 
of the coefficient matrix, which results in the equation
%
%
\begin{equation}
(2p-2-\mu)(2p-\nu)-\lambda^2\mu_+\nu_+=0,
\label{pequation}
\end{equation}
%
%
where we have introduced the notation
%
%
\begin{align*}
\mu \equiv \mu_+\mu_-&=(E-V-b)^2-\Delta^2, \\
\nu \equiv \nu_+\nu_-&=(E-V+b)^2-\Delta^2.
\end{align*} 
%
%
The solutions are
%
%
\begin{equation}
p_\pm(E) \!\!= \!\!\frac{1}{4}\!\!\left[ \mu+\nu+2 \pm \!\!\sqrt{(\mu + 2 -\nu)^2+4\lambda^2\mu_+\nu_+} \right]\,.
\label{eqp}
\end{equation}
%
%
By solving Eq.~\eqref{dcoeff} we then obtain 
%
%
\begin{align}
\psi^{\gtrless}_p(x)& \propto
\begin{pmatrix}
\pm \sqrt{2}(2p-\nu)(p-1)D_{p-2}(\pm q) \\
i\mu_+ (2p-\nu)D_{p-1}(\pm q)\\
\lambda \mu_+\nu_+ D_{p-1}(\pm q)\\
\pm i\sqrt{2}\lambda \mu_+ D_p(\pm q)
\end{pmatrix},
\label{wave>} 
\end{align}
%
%
where $p$ must be replaced by either $p_+$ or $p_-$. 
The set $ \left\{ \psi^>_{p_+},\psi^>_{p_-},\psi^<_{p_+},\psi^<_{p_+} \right\}$ is a basis in the 
space of eigenfunctions of ${\mathcal H}(k)$ with eigenvalue $E$, and thus the general eigenfunction 
can be written as
%
%
\begin{align}
\psi = c_1\psi^>_{p_+} + c_2\psi^>_{p_-} + c_3\psi^<_{p_+} + c_4\psi^<_{p_-} .
\label{gensol}
\end{align}
%
%
For a system homogeneous in  the $x$-direction (with $V(x)=V_0$) the wave functions $\psi^\gtrless_{p_\pm}(x)$
must be normalizable. This happens only if $p_\pm$ are non-negative integers~\cite{Olver:2010}.
In this case, $\psi^<_{p_\pm}$ is proportional to $\psi^>_{p_\pm}$ and can be omitted. 
Then, for each given  $p_+=0,1,2,\dots$ (resp. $p_-=0,1,2,\dots$ )
the solution of Eq.~\eqref{eqp} for $E$ gives a full spectrum of LLs, in 
one-to-one correspondence with spinless graphene's LLs. Therefore, overall 
we obtain two sets of levels, which reduce to the usual twofold spin-degenerate LLs 
when $\lambda=\Delta=b=0$. This spectrum is discussed in detail  in~\cite{DeMartino:2011}. 
Here, we report the results in two simple limits, where explicit analytical expressions can be 
obtained. These results will be useful for comparison with the exact spectrum of the \textit{p-n} junction
discussed in the next Section. 
In the case  $\lambda=0$, omitting the obvious shift by $V_0$, one finds
%
%
\begin{subequations}\label{KMeigen}
\begin{align}
E_{0,\pm} &=  \pm (b - \Delta), \\
E_{n,\pm} &= \pm b +  \text{sign}(n) \sqrt{2|n| +\Delta^2}, \quad n=\pm1,\pm2,\dots 
\end{align}
\end{subequations}
%
%
where the index $\pm$ can be identified with the spin up/down index. 
We note that for $ b=0$ the intrinsic SOI splits the spin degeneracy of the zero-energy states, while 
all other states remain doubly degenerate.
In the case $b=\Delta=0$, one finds the non-zero eigenvalues  given by 
($n=\pm 1,\pm2,\dots$)
%
%
\begin{align} 
E_{n,\pm} =
\text{sign}(n) \left[ 2|n| - 1 +\!\!\frac{\lambda^2}{2} \pm 
\sqrt{\left(1-\frac{\lambda^2}{2}\right)^2 + 2|n|\lambda^2} \right]^\frac{1}{2} ,
\label{RSOeigen}
\end{align}
%
%
in addition to two zero-energy levels~\footnote{For $n=\pm 1$ one only need to keep the solutions $E_{\pm 1,+}$.}.
We observe that the Rashba SOI splits the degeneracy of all levels except the zero energy ones.
We note, in passing, that when $b=\Delta=0$ the Hamiltonian in Eq.~\eqref{modelham} is unitarily 
equivalent to the Hamiltonian of bilayer graphene in Bernal stacking for a single valley~\cite{McCann:2006ev}, with the spin index 
identified with the layer index and the Rashba SOI identified with the interlayer hopping amplitude.  In the limit $|E| \ll \lambda$, 
from Eq.~\eqref{pequation} we obtain $E\approx \pm \frac{2}{\lambda}\sqrt{p(p-1)}$ for the two smallest solutions,
which indeed coincide with the LLs of bilayer graphene in the effective two-band approximation
when $p$ is a non-negative integer~\cite{McCann:2006ev}.


\section{Exact solution of the \textit{p-n} junction}
\label{secIII}

%
%
\begin{figure*}
\centering
\includegraphics[width=0.3\textwidth]{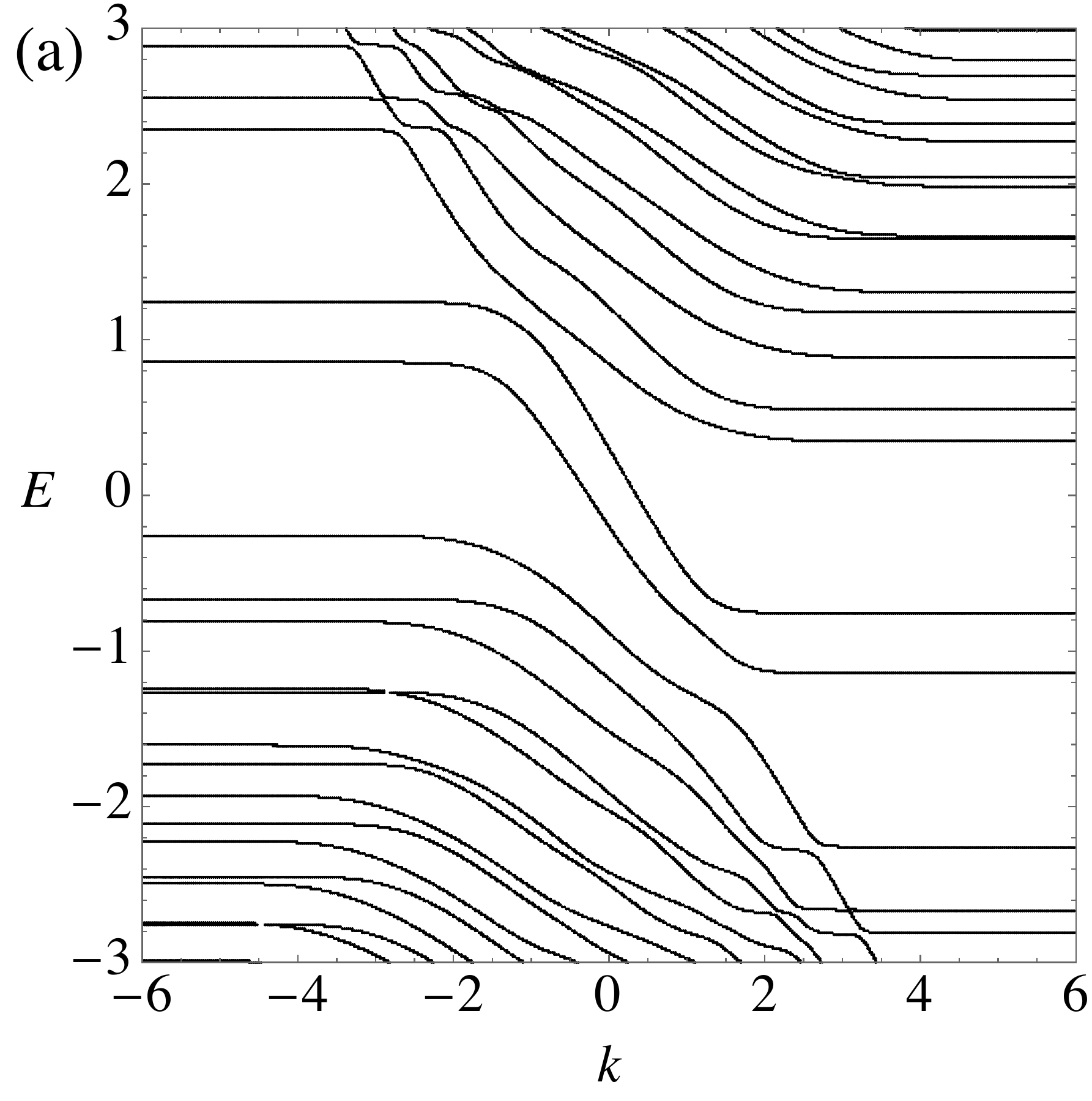}
\includegraphics[width=0.3\textwidth]{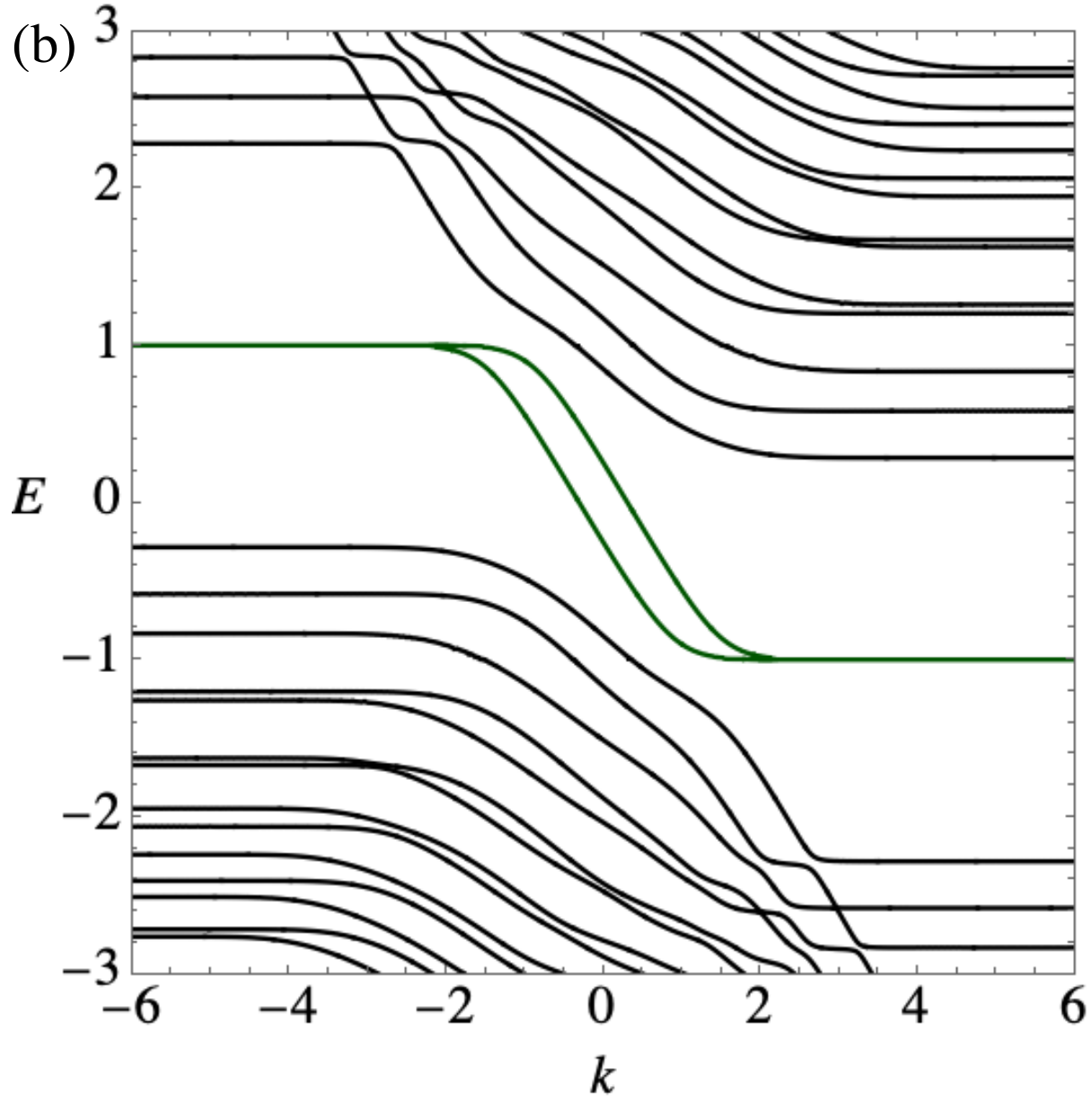}
\includegraphics[width=0.3\textwidth]{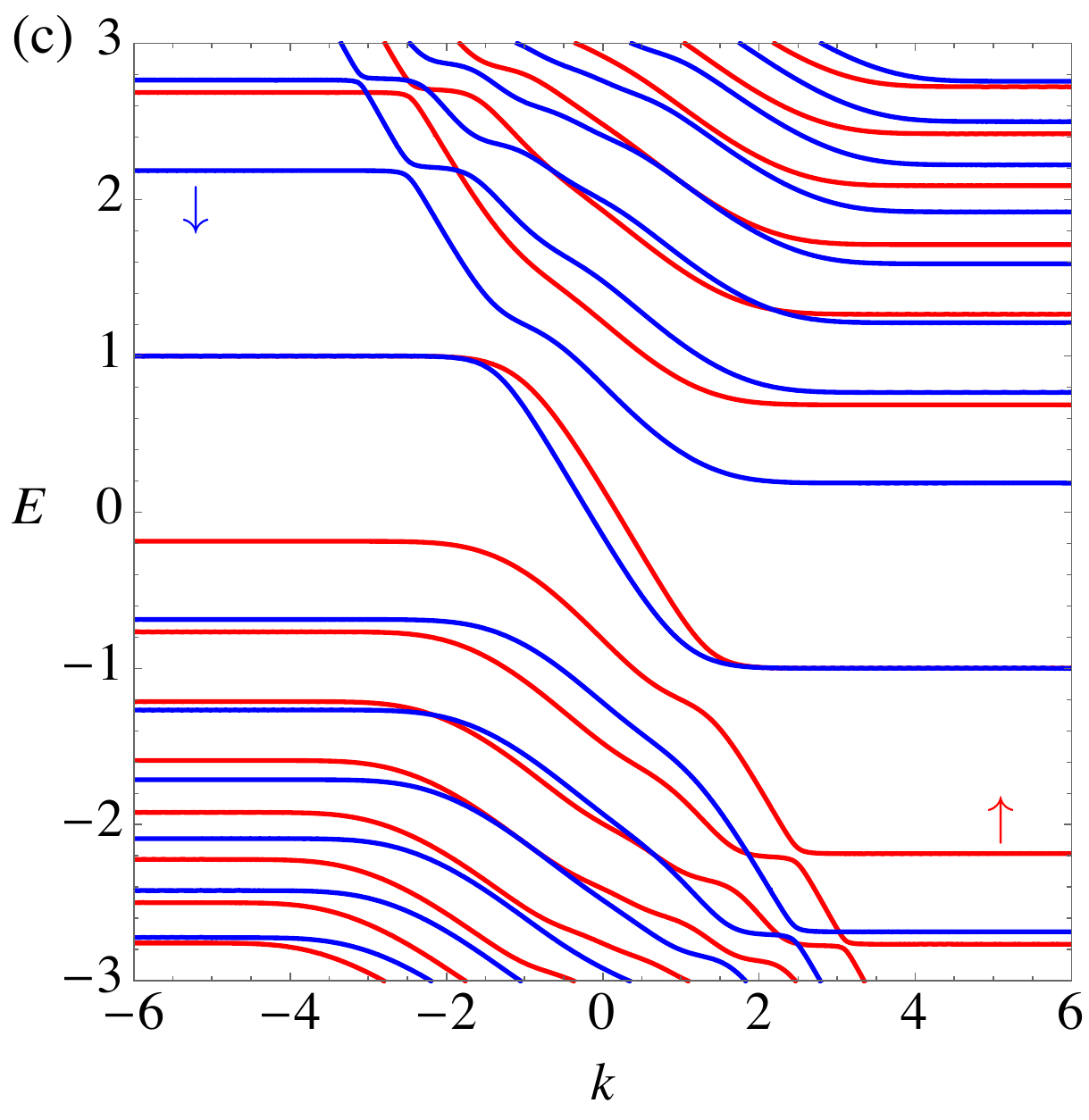}
\caption{Spectra of the \textit{p-n} junction for various parameter sets: (a) 
$b=0.01$, $\Delta=0.25$, and $\lambda=0.7$; (b) $b=\Delta=0$, $\lambda=0.7$,
the green lines mark the zero modes; (c)  $b=\Delta=0.25$, 
the red (resp. blu) lines correspond to up (resp. down) spin projection along $z$. In all the panels we have set $V_0=1$. In
panels (b) and (c) the bulk LLs are given respectively by the expressions in 
Eq.~\eqref{RSOeigen} and \eqref{KMeigen} shifted by $\pm V_0$.}\label{fig2} 
\end{figure*}
%
%
In this section we use the general solution in Eq.~\eqref{gensol} 
to construct the exact eigenstates for the \textit{p-n} junction problem.
In the presence of a non-uniform potential the LLs become dispersive, 
and we shall determine the dispersion relations $E=E(k)$. For the potential step given in 
Eq.~\eqref{step} we need to impose the continuity of the wave function at the interface 
$x=0$ between the $n$-doped region ($x<0$, $V=-V_0$) and the $p$-doped region ($x>0$, $V=V_0$). 
Taking into account the normalizability properties of the parabolic cylinder functions, the general 
form of the wave function is given by 
%
%
\begin{align}
 \psi(x) = 
 \begin{cases}
 c_1\psi^<_{p_+}(x) + c_2\psi^<_{p_-}(x)& x<0,  \\
 c_3\psi^>_{p_+}(x) + c_4\psi^>_{p_-}(x) & x>0,
 \end{cases}
\end{align}
%
%
and the matching condition at $x=0$ reads
%
%
\begin{align}
 c_1\psi^<_{p_+}(0) + c_2\psi^<_{p_-}(0) = c_3\psi^>_{p_+}(0) + c_4\psi^>_{p_-}(0) ,
\end{align}
%
%
where $p_\pm$ are related to the energy $E$ as given in Eq.~\eqref{eqp}.
Therefore we get a homogeneous  linear system
%
%
\begin{align}
{\bf W} {\bf c}={\bf 0},
\label{system}
\end{align}
%
%
with the matrix of coefficients given by 
%
%
\begin{align}
{\bf W}= 
\begin{bmatrix}
\psi^<_{p_+}(0) &  \psi^<_{p_-}(0) &  -\psi^>_{p_+}(0) & -\psi^>_{p_-}(0) 
\end{bmatrix},
\end{align}
%
%
where the spinors $\psi^{\gtrless}_{p_\pm}(0)$ define the columns of the matrix ${\bf W}$,
and
%
%
$$
{\bf c}^T = \begin{pmatrix}  c_1 & c_2 & c_3 & c_4 \end{pmatrix} .
$$
%
%
This linear system admits a non-trivial solution only if the determinant 
vanishes:
%
%
\begin{align}
\det \left[ {\bf W} (k,E) \right] =0 .
\label{spectralequation}
\end{align}
%
%
This condition  implicitly defines the exact spectral branches 
of the \textit{p-n} junction problem.
We study this equation numerically.  For given $k$, we find infinitely many solutions $E=E_{n,\pm}(k)$
that can be labelled by a Landau level index $n\in {\mathbb Z}$ and an index $\pm$ which, 
in the limit of vanishing $\lambda$, reduces to a spin projection quantum number~\footnote{
We note that, due to the spin degeneracy of all LLs at $\lambda=\Delta=b=0$, the limits $\lambda\rightarrow 0$ 
and $(b,\Delta) \rightarrow 0$ do not commute:  the limit $\lambda\rightarrow 0$ at finite $b$ or $\Delta$ 
produces eigenstates of $s_z$, whereas  the limit $\lambda\rightarrow 0$ at $b=\Delta=0$ gives eigenstates of $s_x$.}.
Then, for any given solution $\left\{ k, E_{n,\pm}(k) \right\}$ of Eq.~\eqref{spectralequation},
we solve the linear system~\eqref{system} to obtain the corresponding eigenfunction, 
which can then be normalized.
Let us now discuss in detail our findings.

The spectrum consists of infinitely many branches of dispersive LLs.
This is illustrated in Fig.~\ref{fig2}a for a generic case, with $\lambda$, $\Delta$, and $b$ all finite. 
Two special cases of particular interest are illustrated in Fig.~\ref{fig2}b, where $b=\Delta=0$, and $\lambda$ is finite; 
and  in Fig.~\ref{fig2}c, where $\lambda=0$  and $\Delta$ and $b$ are  finite and equal. 
 
The flat portions of each level (for $|k|\gg1$) in Fig.~\ref{fig2} correspond to bulk states.
The approximately linearly dispersing portions visible throughout the spectrum 
correspond to states mainly localized close to the interface. Due to the Rashba SOI, which 
mixes spin-up and spin-down, these levels have  $k$-dependent spin polarization, 
which we discuss in Sec.~\ref{secIV} below. Adjacent levels generally exhibit anticrossings, see Figs.~\ref{fig2}a and~\ref{fig2}b,
unless $\lambda=0$, in which case the projection of spin along the $z$-direction is a good quantum number, and one finds
level crossings instead, as in Fig.~\ref{fig2}c. 

For a not too large amplitude of the potential step, there exists an energy window
around zero energy in which there is a single pair of interface modes. This is the situation
illustrated in the three panels of Fig.~\ref{fig2}.  If the Fermi level lies in this energy window, the observable 
properties of the system will be determined by these linearly dispersing chiral zero modes.
In the rest of the paper we will mainly focus on them. As they evolve from the two zero-energy modes 
of the uniform system, we will refer to them as  the zero modes of the \textit{p-n} junction. 
Close to zero energy their dispersion is linear to a very good approximation, and 
they can be regarded as the quantum states corresponding to spin-resolved 
quasiclassical snake states. 

Let us consider now the case $b=\Delta=0$, illustrated in Fig.~\ref{fig2}b.
Remarkably, the dispersions of the two zero modes exhibit a relative shift in wave vector reminiscent of the
horizontal shift between the energy dispersions in a single-channel quantum wire with Rashba SOI \cite{Datta:1990,Bercioux:2015}.
The modes become again degenerate (with exponential accuracy) for $|k|\rightarrow \infty$.
We can rationalize this behaviour as follows. In the gauge we are working in, 
the electronic states can be visualized as stripes along the $y$-direction, localized in the $x$-direction 
around $x=-k$ within a distance of the order of the magnetic length.
For $|k|$ very large, their centre is far from the interface and they are not affected by the potential step. 
The solution for the uniform case  then shows that there are two degenerate zero modes, whose energies 
do not depend on the Rashba coupling. This occurs because they both have the same vanishing (or more precisely, 
exponentially small) sublattice component. 
Since the Rashba SOI couples the amplitudes of electrons with opposite spin on different sublattices, its matrix element 
will be at most exponentially small, and it will not lift the spin degeneracy. 
For $|k|\approx 0$, instead,  the electronic states are close to the interface and the zero modes have a 
significant amplitude on both sublattices. Then the Rashba SOI couples them, lifting the degeneracy.

\begin{figure}
\centering
\includegraphics[width=0.45\textwidth]{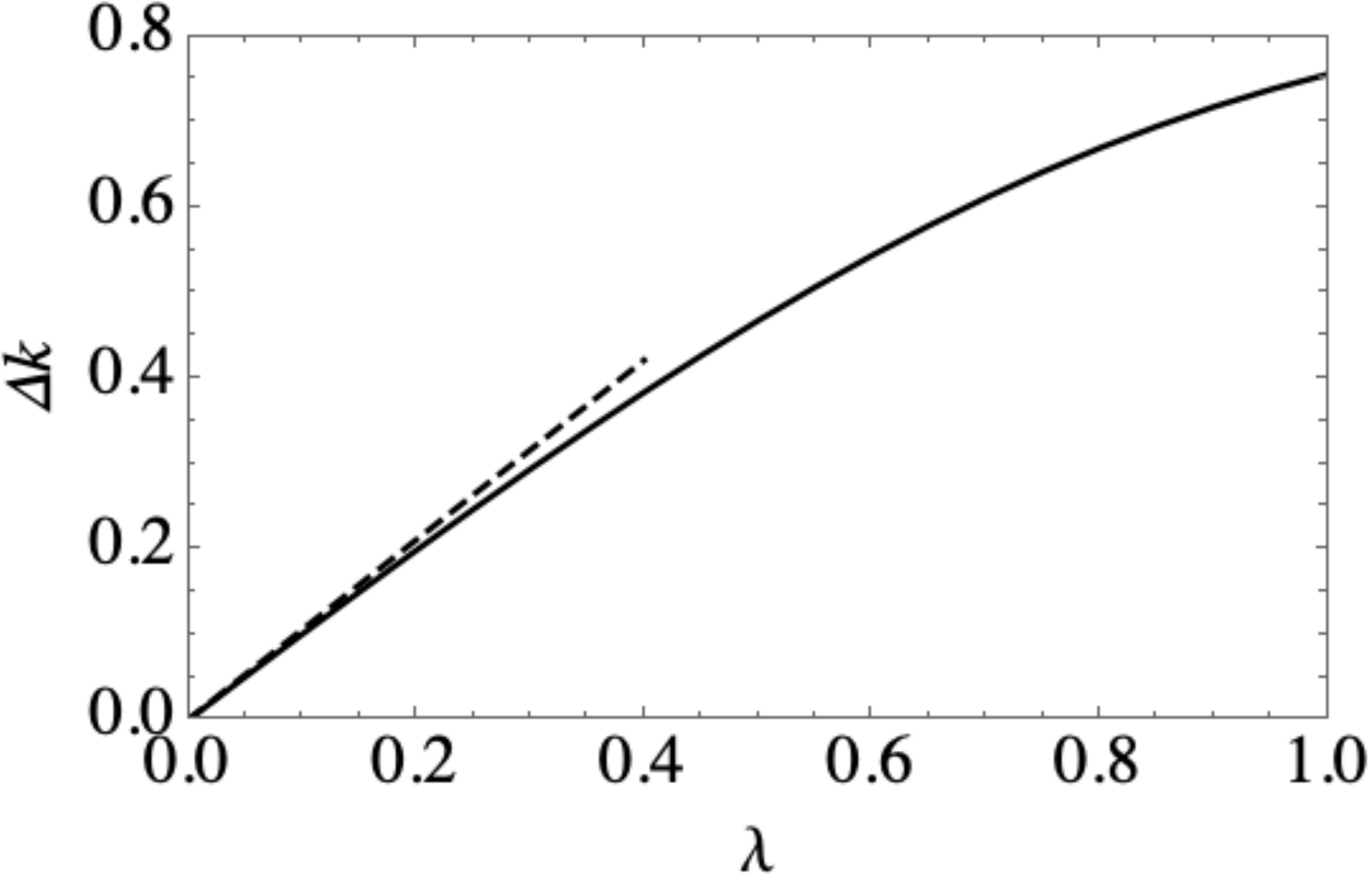}
\caption{
Plot of the zero-mode wave vector splitting $\Delta k$ at $E=0$ versus  the Rashba SOI. (Here we set $\Delta=b=0$ and $V_0=1$.)
The solid line is the exact result, the dashed line the perturbative result in Eq.~\eqref{perturbative}.}
\label{fig3}
\end{figure}

The wave vector splitting between the two zero modes can be calculated analytically
in perturbation theory in $\lambda$ (see App.~\ref{appA}). We find the 
simple result
%
%
\begin{align}
\Delta k \equiv k_+-k_-= \frac{\lambda}{\hbar v_\text{F}},
\label{perturbative}
\end{align}
%
%
where $k_\pm$ are defined as the solutions of $E_{0,\pm}(k)=E$, and we have
reintroduced the units. Interestingly, the splitting is independent of energy, of magnetic field and even of the 
potential difference across the junction.
Using our exact dispersions, we have also computed  the exact splitting as a function of $\lambda$ numerically.
The exact result is compared with the perturbative one in Fig.~\ref{fig3}, which shows that the two are in good agreement
up to $\lambda \approx 0.2$.

The \textit{p-n} junction then hosts at given energy  two degenerate modes at different wave vector with
finite overlap, and we expect interference effects in the observables induced by the Rashba SOI, which we discuss in the next section.  
A relative shift of the zero-mode dispersions can be observed also for vanishing $\lambda$ and finite $b$ and $\Delta$.
In this case, however, the degeneracy is restored at large $k$ only in the non generic situation that $b=\Delta$
illustrated in Fig.~\ref{fig2}c. More importantly, for $\lambda=0$ the two modes are eigenstates of $s_z$, thus 
their overlap vanishes and interference effects are not possible. 
%
%
%
%


\section{Observables}
\label{secIV}

From now on, we focus solely on the effects of the Rashba SOI ($\Delta=b=0$)
and we restrict to an energy window in which there exist only the two dispersive zero modes.  
At given energy $E$, the electronic state is then a superposition of 
snake states at two different wave vectors
%
%
\begin{align}
\Psi_{E}(x,y) \propto A_1 e^{i k_+y} \psi_{k_+}(x)  +  A_2 e^{i k_-y} \psi_{k_-}(x).
\end{align}
%
%
All observables, bilinear in the wave function, will contain a term 
uniform in $y$ and an interference term oscillating along $y$, with period given by $2\pi\Delta k^{-1}$.  
These  oscillating terms manifest themselves in correlation functions, 
as we discuss below, provided the overlap between the states at $k_\pm$ does not vanish. This happens only if
the Rashba SOI is different from zero. If $\lambda=0$, the two modes are spin eigenstates with opposite spin projection 
and hence are orthogonal. As a consequence, the presence of such oscillating terms in correlation functions
is a hallmark of Rashba SOI and provides a way of measuring the strength of the coupling.
%
%
\begin{figure}
\centering
\includegraphics[width=0.4\textwidth]{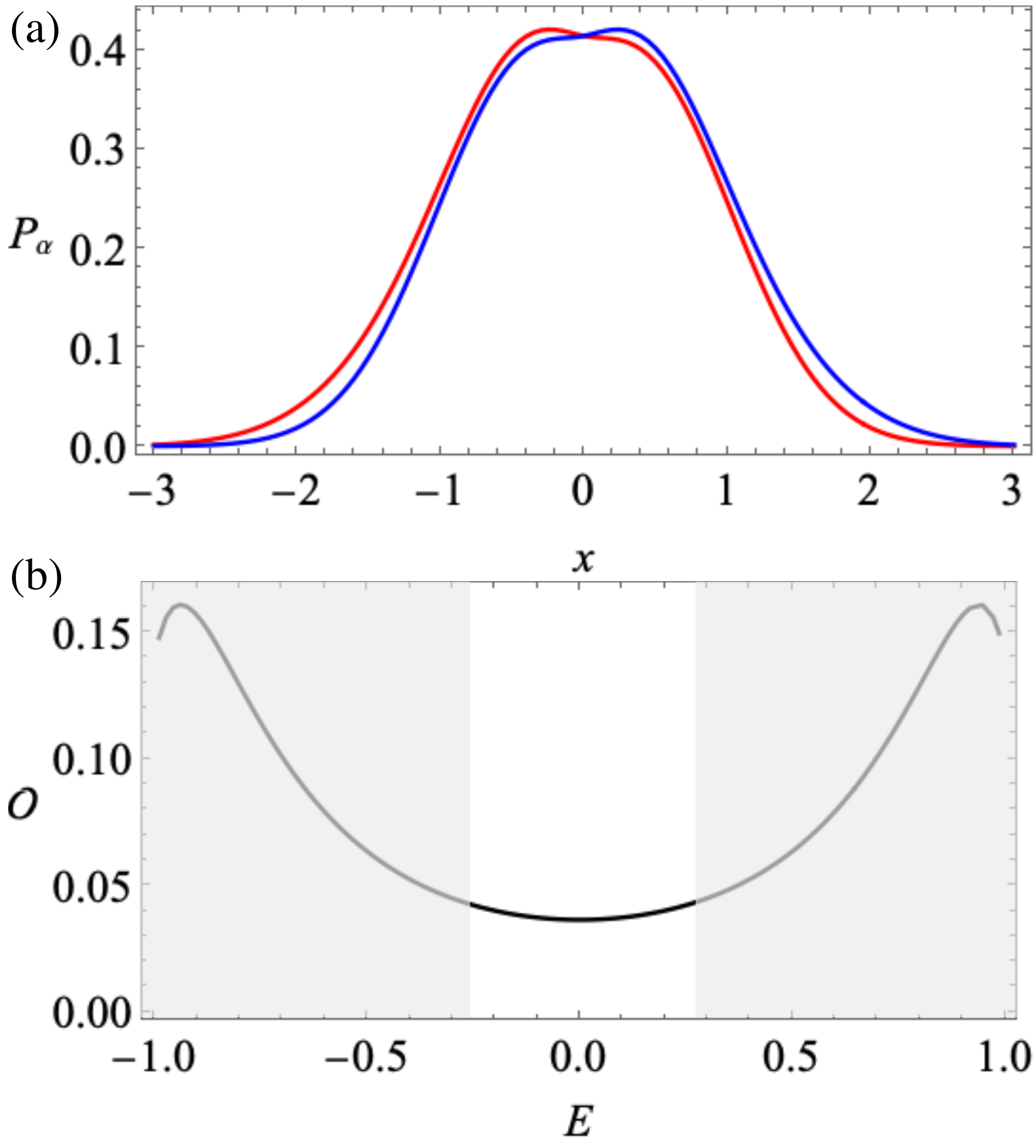}
\caption{(a) Probability density profile for the two states at $E=0$. 
(b) Energy dependence of the overlap of the two zero modes. The grey areas indicate energy windows where 
there are additional states in the spectrum on top of the two zero modes (at different wave vectors).  Here we set $V_0=1$ 
$\lambda=0.7$ and $b=\Delta=0$.}
\label{fig4}
\end{figure}
%
%
%
%
\begin{figure}
\centering
\includegraphics[width=0.45\textwidth]{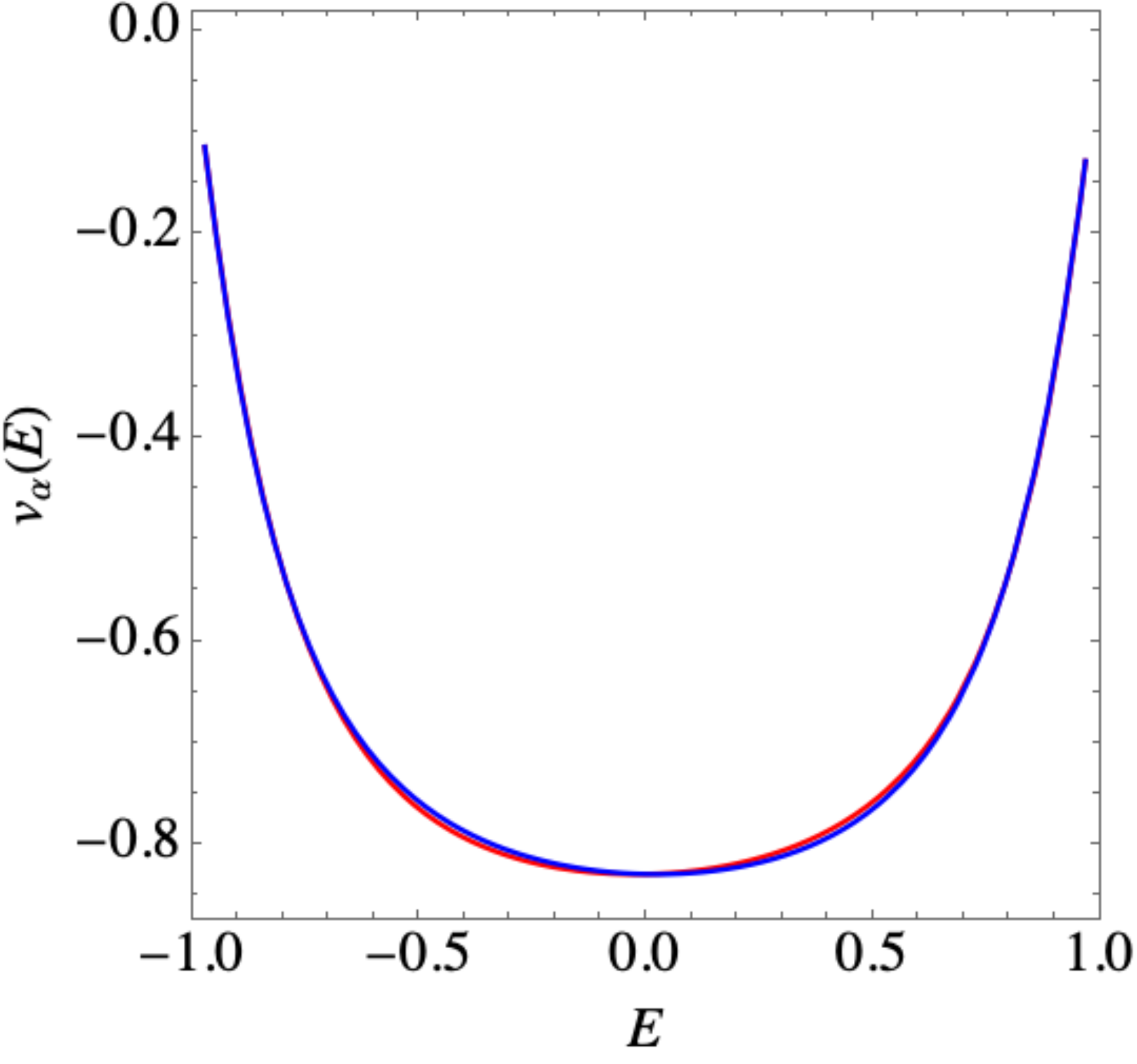}
\caption{The group velocities $v_\alpha$ ($\alpha=\pm$) of the two zero modes
as a function of energy for $V_0=1$, $\lambda=0.7$ and $b=\Delta=0$.}
\label{fig5} 
\end{figure}
%
%

%
%
\begin{figure*}[!htb]
\centering
\includegraphics[width=0.355\textwidth]{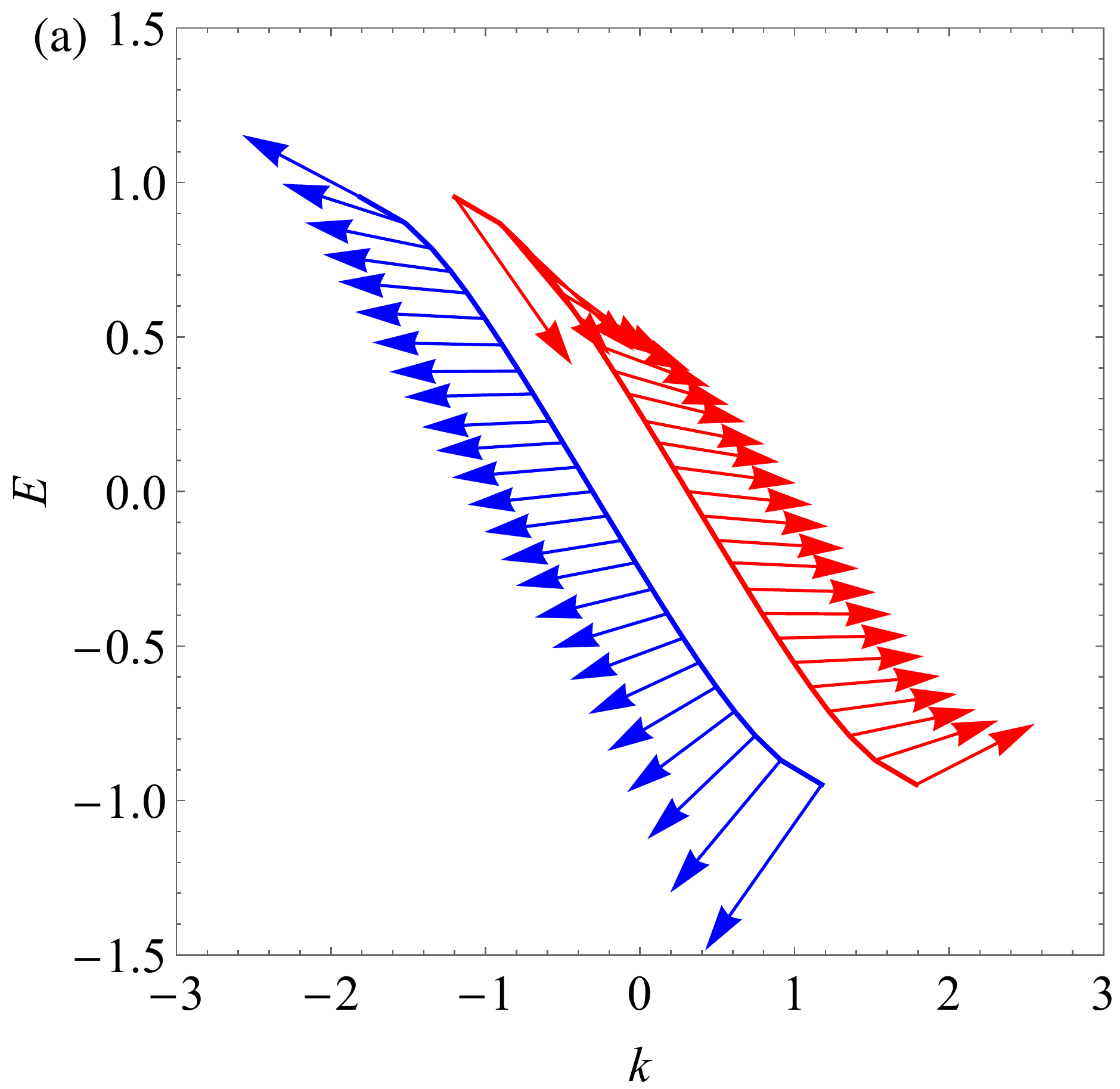}
\includegraphics[width=0.35\textwidth]{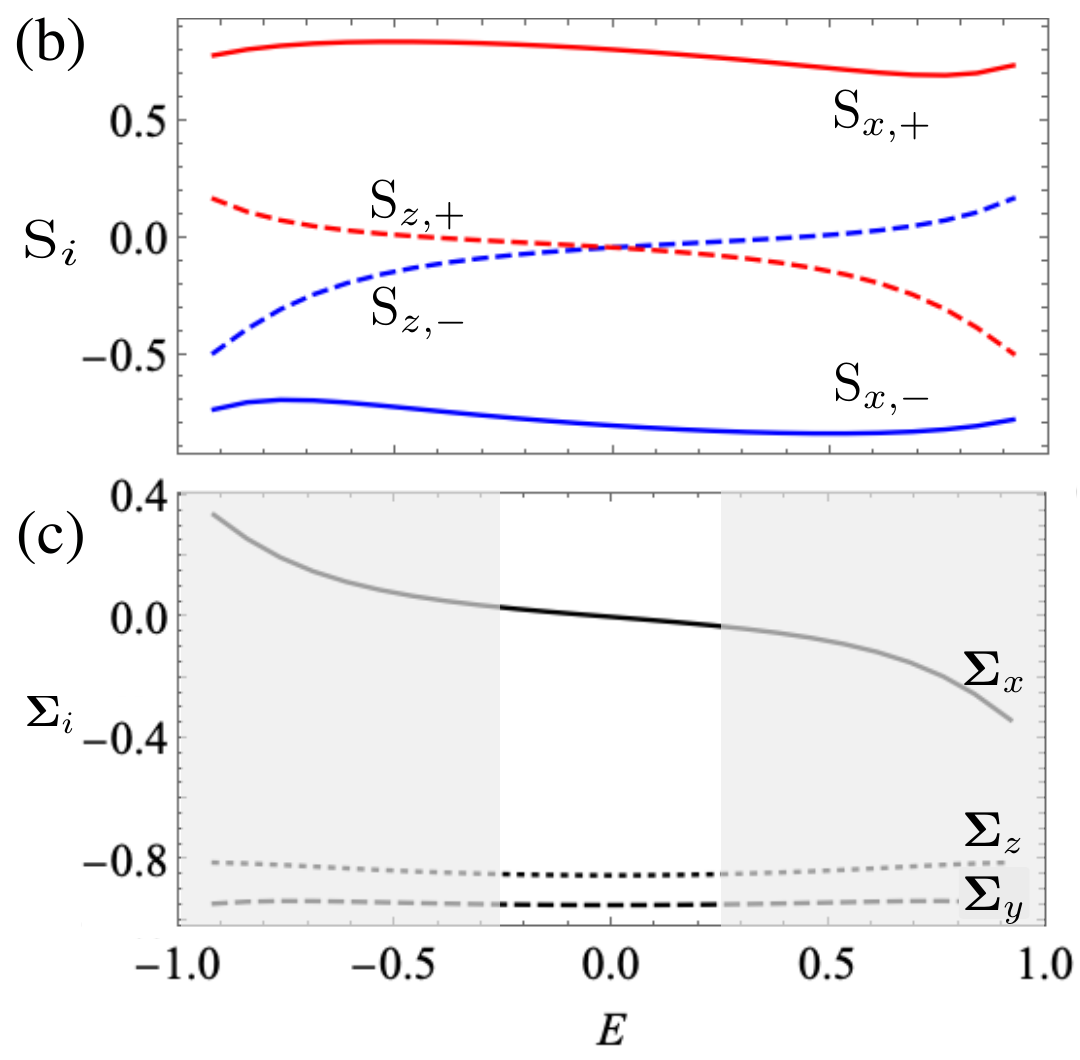}
\caption{(a) The spin polarization ${\bf S}_\pm$ as function of $E$ and $k$ for the two zero modes. 
Since $S_y=0$, we plot only the $x$ and $z$ components.
(b) Projections of the spin polarizations as function of $E$.
(c) Components of the matrix element of the spin operator between the zero modes, see Eq.~\eqref{spinstaggered}. 
(For $\Sigma_{y}$, which is imaginary, we actually plot $-i\Sigma_y$.) 
The grey areas indicate energy windows where there are additional states in the spectrum on top of the two zero modes
(at different wave vectors).
In all the panels we have used $V_0=1$, $\lambda=0.7$ and $\Delta=b=0$.}
\label{fig6}
\end{figure*}
%
%
The probability density associated to the states at $k_\pm$, $P_\alpha(x) =\psi^\dagger_{k_\alpha}(x)\psi_{k_\alpha}(x)$ ($\alpha=\pm$)
is illustrated in Fig.~\ref{fig4}a for the states at $E=0$. One observes that the profiles are localized at $x=0$ within 
a magnetic length, are skewed, with maxima on opposite sides of the interface. 
Although the wave functions seem to have a substantial spatial overlap,  the overlap integral
%
%
\begin{align}
{\cal O} = \langle \psi_{k_+} |  \psi_{k_-} \rangle  =  \int dx \,  \psi^\dagger_{k_+}(x) \psi_{k_-} (x),
\label{overlap}
\end{align}
%
%
(which turns out to be real) is rather small, as shown in Fig.~\ref{fig4}b. 
It reaches a minimum at $E=0$,  but remains finite in the whole energy window.  
This can be rationalised with the observation that a finite Rashba SOI produces a spin texture in 
the eigenfunctions, which would otherwise have opposite spin orientations and would be orthogonal.     
As discussed, above, a finite overlap is important as it is the weight of the oscillating term in 
the density-density correlator. 

Next, we consider the mode velocities. The velocity operator is
${\bf v} = s_0 \otimes {\bm \sigma}$, and  the group velocities of the two zero 
modes are given by 
 %
%
\begin{equation}\label{velocity}
{\bf v}_\pm =\langle \psi_{k_\pm}  | s_0\otimes {\bm  \sigma}  | \psi_{k_\pm} \rangle.
\end{equation}
%
%
We find $v_{\pm,x}=0$, while the $y$-components $v_{\pm,y}$, which coincide with the slopes of the 
zero-mode dispersions $E'_{0,\pm}(k_\pm)$, are illustrated in Fig.~\ref{fig5} as function of energy.
In the whole energy range the velocities practically coincide. 
They are finite and  reach their maximum (in absolute value) at $E=0$. Close to $E=0$ they 
are almost constant, which implies that the dispersions are almost perfectly linear.
We have checked numerically that they depend only weakly on $\lambda$. 
This holds also in the limit  $V_0\ll \lambda$.
In fact, a straightforward perturbative calculation in $V_0$ (but exact in $\lambda$)
shows that for $k\approx  0$ the velocities can be expressed as 
%
%
\begin{align}
v_{\pm,y} = - C(\lambda/\hbar \omega_c)  \frac{V_0\ell_B}{\hbar},
\end{align}
%
%
where $C(x) =  \frac{2+x^2/2}{\sqrt{\pi}(1+x^2/2)}$
and  we have restored units. This value does not depend on graphene's Fermi velocity. 
Since $V_0/e\ell_B$ can be interpreted as the electric field $E_x$ 
induced by the potential step in the $x$-direction and $\ell^2_B \propto 1/B$, $V_0\ell_B/\hbar$ 
corresponds to the classical drift velocity of a charged particle in crossed magnetic and electric fields, $cE_x/B$
\cite{Cohnitz:2016}.
The effect of the Rashba SOI is a renormalization factor which decreases for increasing $\lambda$ from $2/\sqrt{\pi}$ (for $\lambda=0$)
to $\frac{1}{\sqrt{\pi}}$ (for $\lambda \gg \hbar \omega_c$) and thus tends to slightly reduce the velocities.

Next, we consider the spin polarization. The spin operator is given by
${\bf S} = {\bf s} \otimes  \sigma_0$
 (for simplicity we omit the factor $1/2$), so the expectation value of the spin 
for the two zero modes is given by 
%
%
 \begin{align}
{\bf S}_\pm =\langle \psi_{k_\pm}  | {\bf s} \otimes  \sigma_0 | \psi_{k_\pm} \rangle.
\label{spinexpectation}
 \end{align}
%
%
We find that $S_{y,\pm}=0$ and that ${\bf S}_{\pm}$ points predominantly in the $x$-direction.
This can be rationalized as due to the fact that the Rashba SOI induces a spin precession in a 
direction perpendicular to the direction of motion and, in the geometry considered here, it selects 
the eigenstates of $s_x$  (see discussion at the end of App. \ref{appA}). The $z$-component switches 
sign across $E=0$ and we observe the symmetry properties $S_{z,-}(E) = S_{z,+}(-E)$ and $S_{x,-}(E) = - S_{x,+}(-E)$.

The matrix element of the spin operator between the two zero modes is given by  
%
%
\begin{align}
{\bf \Sigma}_{\pm} & = \langle \psi_{k_\pm}  |   {\bf s} \otimes  \sigma_0 |  \psi_{k_\mp} \rangle, 
\label{spinstaggered}
\end{align}
%
%
and is illustrated in Fig.~\ref{fig6}. It plays an important role in the spin-spin correlation function, 
because it is the weight of the interference term. We find that $\Sigma_{x,\pm}$ and $\Sigma_{z,\pm}$ are real, with
 $\Sigma_{x(z),+}=\Sigma_{x(z),-}$, while $\Sigma_{y,\pm}$ are imaginary, with  $\Sigma_{y,+}=-\Sigma_{y,-}$.

We turn now to the density-density and spin-spin correlations. While we have access  in principle to the full exact 
correlators, for the purpose of a study of the low-energy properties around $E=0$, it is sufficient to focus on the 
linear part of the spectrum and describe our system as a pair of one-dimensional non-interacting chiral electronic modes, 
with (second-quantized) Hamiltonian given by
%
%
\begin{align}
H & \approx  \sum_{\alpha=\pm}  v_\alpha \int dy   L^\dagger_{\alpha} (-i \partial_y) L_\alpha,
\end{align}
%
%
and the fermion field operator given by
%
%
\begin{align}
\Psi (x,y) & \approx \sum_{\alpha=\pm} e^{ik_\alpha y} \psi_{\alpha}(x)   L_\alpha(y),
\label{1dfermion}
\end{align}
%
%
where $\psi_\alpha(x)$ are the exact eigenstates of the zero modes at $E=0$ and $L_\alpha(y)$ fermionic
operators with usual anti-commutation relations:
%
%
$$
\left\{ L_\alpha(y), L^\dagger_\beta(y')\right\} =\delta_{\alpha\beta}\delta(y-y'). 
$$
%
%
Using the expression for the field operator \eqref{1dfermion}, the effective one-dimensional 
density operator, obtained by integrating the two-dimensional density operator over $x$,  is given by
%
%
\begin{align}
\rho (y)&=\int dx\, \Psi^\dagger (x,y)\Psi (x,y) \nonumber \\
& \approx \sum_{\alpha=\pm} \rho_\alpha (y) + 
 \sum_{\alpha=\pm} {\cal O} 
e^{-i\alpha \Delta k \, y }  n_\alpha (y),
\end{align}
%
%
with 
%
%
$$
\rho_\alpha(y) =L^\dagger_\alpha  L_{ \alpha }(y), \quad n_\alpha (y)= L^\dagger_\alpha  L_{\bar \alpha }(y),
$$
%
%
and the overlap integral ${\cal O}$, defined in Eq.~\eqref{overlap}, is evaluated at $E=0$. 
Therefore, the density-density correlator will contain two contributions:
%
%
$$
\langle \rho(y) \rho(0) \rangle = \sum_{\alpha}  \langle \rho_\alpha (y) \rho_\alpha (0) \rangle
+ \sum_\alpha e^{-i \alpha \Delta k \, y } {\cal O}^2\langle n_\alpha (y) n_{\bar \alpha} \rangle, 
$$
%
%
The uniform term is the usual correlator of one-dimensional chiral fermions:
%
%
\begin{align}
 \sum_{\alpha}  \langle \rho_\alpha (y) \rho_\alpha (0) \rangle  
 = \frac{1}{4\pi^2y^2}.
 \end{align}
%
%
The interference term is given by
%
%
\begin{align}
& \sum_{\alpha} {\cal O}^2  e^{-i\alpha \Delta k \, y} \langle n_\alpha (y) n_{\bar \alpha} (0) \rangle 
 =\frac{  {\cal O}^2\cos (\Delta k \, y)}{4\pi^2y^2}.
 \end{align}
%
%
As anticipated, the density-density correlator exhibits spatial oscillations with a period proportional to the inverse of $\Delta k$.
The structure factor, which is essentially its Fourier transform,  will then exhibit a pronounced observable feature at $q=\Delta k$.

The spin-spin correlator has a similar structure.
The effective one-dimensional spin-density operator  is given by
%
%
\begin{align}
S(y) & =\int dx \, \Psi^\dagger (x,y) {\bf  s}\Psi (x,y) \nonumber \\
&\approx  \sum_{\alpha} {\bf S}_\alpha \rho_\alpha (y) +  \sum_\alpha 
e^{-i \alpha \Delta k\, y } {\bf \Sigma}_{\alpha} n_\alpha (y),
\end{align}
%
%
where the matrix elements ${\bf S}_\alpha$ and ${\bf \Sigma}_\alpha$, defined in Eqs. \eqref{spinexpectation} 
and \eqref{spinstaggered}, are evaluated at $E=0$. The spin-spin correlator then will be 
%
%
\begin{align}
\langle S^i(y) S^j(0) \rangle &= \sum_\alpha  {S_\alpha^i}{S_\alpha^j} \langle \rho_\alpha(y)  \rho_\alpha(0) \rangle  \nonumber \\
&+\sum_\alpha {\Sigma_\alpha^i}{\Sigma_{\bar \alpha}^j}   e^{-i \alpha \Delta k\,  y} \langle n_\alpha(y)  n_{\bar \alpha}(0) \rangle.
\end{align}
%
%
Therefore the spin structure factor will present a feature at momentum $q=\Delta k$,
indicating the existence of an interference term.

\section{Conclusions and Outlook}
\label{secV}
 
In this paper we have studied the effects of finite spin-orbit interaction on the 
low-energy electronic properties of \textit{p-n} junctions in graphene subject to a 
perpendicular magnetic field. We have found the exact solution of the appropriate Dirac-Weyl 
equation and thereby determined the exact spectrum of dispersive LLs. 
The spectrum contains a pair of propagating modes which cross zero energy with approximately 
linear dispersion, and can be interpreted as quantum limit of snake states. 
When the Fermi level lies close to $E=0$ between two bulk LLs, the transport properties 
are dominated by these snake states, similar to those forming in usual graphene \textit{p-n} junctions in 
magnetic field. However, the Rashba SOI  induces a mismatch in the Fermi 
momenta of these states, which has observable consequences. In fact, we predict that the 
density-density and the spin-spin correlation functions have a spatially oscillating behaviour 
along the junction, which could be tested experimentally.

The relative horizontal shift of the dispersions of the interface zero modes and the $k$-dependence of their 
spin polarization are strongly reminiscent of the analogous effects in a single-channel quantum wire 
with Rashba SOI. This analogy suggests that the \textit{p-n} junction works as a Datta-Das channel. 
If one injects electrons with definite spin polarization at some point along the junction, and measure correlations 
downstream, in the presence of a substantial Rashba SOI one should observe oscillations due to interference effects 
between the two modes. The junction could then provide a realization of a spin-FET. With respect to the 
case of a quantum wire, this realization would have the advantage that the system is chiral, and 
therefore it is practically immune to backscattering by impurities. Moreover, although one would expect, 
in principle, large effects of electron-electron interactions on these effectively one-dimensional electronic states,
the chirality of the system suppresses the backscattering processes and prevents the opening of a gap 
at the Fermi level. The e-e interactions might then only renormalize the velocities and modify the exponents in 
the power law decay of correlations, but could not spoil the effects discussed here. We leave the detailed study of 
the e-e interaction effects for future works. 

We hope that our work will stimulate further experimental 
and theoretical activity on the properties of snake states in graphene's \textit{p-n} junction, and 
that the predictions put forward here can be soon tested experimentally.

\acknowledgments
We thank Reinhold Egger and Tineke van den Berg for a careful reading of the manuscript and useful suggestions. 
The work of DB is supported by Spanish Ministerio de Ciencia, Innovation y Universidades (MICINN) 
under the project FIS2017-82804-P, and by the Transnational Common Laboratory \textit{Quantum-ChemPhys}.

\appendix

\section{Perturbation theory}
\label{appA}

In this appendix we briefly discuss the calculation of the spectrum of the Hamiltonian in Eq.~\eqref{ham} 
by perturbation theory  in $\lambda$. We use this result to find an explicit formula for 
the dependence of the splitting $\Delta k$ on the Rashba SOI. 

For $\lambda=0$ the Hamiltonian \eqref{ham} is 
block-diagonal in spin. Each block coincides with the Hamiltonian studied in~\cite{Cohnitz:2016} (up to the intrinsic 
SOI term and the Zeeman term, which can be easily incorporated) and we can borrow their results.
The wave functions can be expressed as 
%
%
\begin{align*}
\psi_{k,s}(x) = |s\rangle\otimes\phi_{k,s}(x), \quad s=\uparrow/\downarrow,
\end{align*}
%
%
where $|s\rangle $ are the eigenstates of $s_z$ and
%
%
\begin{align}
\phi_{k,\uparrow}(x)= \frac{D_{p(-x)}(-\epsilon(x)\sqrt{2}k)}{\sqrt{\cal N}}  
\begin{pmatrix}
\frac{\epsilon(x) \mu_-(x)}{\sqrt{2}} D_{p(x)-1}(\epsilon(x) q) \\
i D_{p(x)}(\epsilon(x) q) 
\end{pmatrix}.
\label{continuous}
\end{align}
%
%
Here $\epsilon(x) =\text{sign}(x)$, ${\cal N}$ is a normalization constant and 
%
%
\begin{align*}
\mu_-(x) & = E-V(x)-b +\Delta,\\
p(x) & =\frac{1}{2} \left[ (E-V(x)-b)^2-\Delta^2 \right].
\end{align*}
%
%
The wave function \eqref{continuous}
is continuous at $x=0$ provided the matching condition
%
%
\begin{align*}
\det \begin{bmatrix} 
\phi_{k,\uparrow}(0^-) & - \phi_{k,\uparrow}(0^+)
\end{bmatrix}=0
\end{align*}
%
%
is satisfied. This condition determines the eigenvalues $E$ and plugging the eigenvalues 
back into Eq.~\eqref{continuous} one obtains the corresponding eigenstates $\phi_{k,\uparrow}$ . 
The eigenstates $\phi_{k,\downarrow}$ and the corresponding eigenvalues are 
obtained by replacing $\Delta$ and $b$ with $-\Delta$ and $-b$.

Henceforth, we  set  $\Delta=b=0$. Then $\phi_{k,\uparrow}=\phi_{k,\downarrow}$
and we can omit the spin index. All eigenstates (at given $k$) are doubly degenerate. 
We focus on the two zero modes, with unperurbed energy $E_0(k)$. 
Using standard degenerate perturbation theory, 
we find that the first order perturbative correction to the energy due to the Rashba SOI 
is  given by
%
%
$$
\delta E_{0,\pm}(k) =\pm h(k),
$$
%
%
where $h(x)$ is the matrix element of the perturbing operator
%
%
\begin{align}
h(k)&= \langle \psi_{k,\uparrow}  | {\cal H}_\text{RSO}| \psi_{k,\downarrow} \rangle \nonumber\\
&=- \frac{\lambda}{{\cal N}} \int dx \, \epsilon(x) \, [D_{p(-x)}(-\epsilon(x)\sqrt{2}k)]^2\times \nonumber  \\
&  \times \frac{(E-V(x))}{\sqrt{2}} D_{p(x)-1}(\epsilon(x) q) D_{p(x)}(\epsilon(x) q) .
\label{h(k)}
\end{align}
%
%
Up to the factor $-\lambda/2$, this  expression is nothing else than the $y$-component of
the group velocity $v^{(0)}_y(k)$ of the unperturbed states.
Therefore, $h(k)=-\frac{\lambda}{2} v^{(0)}_y(k)$ and  the mode dispersions read 
%
%
$$
E_{0,\pm}(k) = E_0(k) \mp \frac{\lambda}{2} v^{(0)}_y(k) + {\cal O}(\lambda^2).
$$
%
%
The difference in wave vector between the two modes at energy $E$
is then simply given by 
%
%
$$
\Delta k \equiv k_+ - k_- = \lambda,
$$ 
%
%
where $k_\pm$ are defined by the equation $E_{0,\pm}(k_\pm)=E$. 
In other words, $\lambda$ produces a relative horizontal shift of the dispersions. 
Remarkably, at this order in $\lambda$,
the shift is independent of the energy, the potential strength, and the magnetic field.
The unperturbed eigenstates that diagonalize ${\mathcal H}_{\text{RSO}}$ 
are the eigenstates of $s_x$:
%
%
\begin{align}
\psi_{k,\pm}(x) =\frac{1}{\sqrt{2}}  (|\uparrow \rangle \pm | \downarrow \rangle)\otimes  \phi_k(x).
\end{align}
%
%
This is consistent with the fact that our exact eigenstates have spin polarization predominantly along 
the $x$-direction, see Fig.~\ref{fig6}, and explains the observation that at $b=\Delta=0$ 
they reduce to eigenstates of $s_x$   in the limit  $\lambda\rightarrow 0$.

\bibliography{biblio}

\end{document}